\documentclass[twocolumn,amsmath,amssymb]{revtex4}

\usepackage{graphicx}
\usepackage{dcolumn}
\usepackage{bm}

\usepackage[all,import]{xy}
\usepackage{warmread}

\def\nhat{{\bf n}}
\def\rotfnc{\chi}

\def\boldgrad{\bm\nabla}
\def\grad{\nabla}

\begin{document}

\title{Dynamics of immersed molecules in superfluids}
\author{Michael J. Quist}
\email{mjq1@cornell.edu}
\author{Veit Elser}

\affiliation{Department of Physics, Cornell University, Ithaca, New York 14850}

\date{\today}

\begin{abstract}
The dynamics of a molecule immersed in a superfluid medium are considered.
Results are derived using a classical hydrodynamic approach followed by canonical
quantization.  The classical model, a rigid body immersed in incompressible fluid,
permits a thorough analysis; its effective Hamiltonian generalizes the usual rigid-rotor
Hamiltonian.  In contrast to the free rigid rotor, the immersed body is shown to have
chaotic dynamics.  Quantization of the classical model leads to new and experimentally
verifiable features.  It is shown, for instance, that chiral molecules can behave as "quantum
propellers": the rotational-translational coupling induced by the superfluid leads to a
nonzero linear momentum in the ground state.  Hydrogen peroxide is a strong candidate
for experimental detection of this effect.  The signature is a characteristic splitting of
rotational absorption lines.  The $1_{01} \rightarrow 1_{10}$ line in hydrogen peroxide,
for example, is predicted to split into three lines separated by as much as $0.01$\;cm$^{-1}$,
which is about the experimental linewidth.
\end{abstract}

\maketitle

\section{Introduction}
\label{secintro}

The dynamics of a molecule moving within a bulk superfluid medium are not
expected to be too different from its dynamics in vacuum.  Like the
vacuum, an unbounded fluid is homogeneous and invariant with respect
to continuous translations and rotations.  Moreover, in the exotic
case of superfluids, the very low level of excitation achieved at low
temperatures implies that there is a vanishing set of modes with which
the immersed molecule can interact.  These expectations are borne out
in the spectroscopy of single impurity molecules in $^{4}$He
nanodroplets, where the superfluid medium has little effect on the
rotational spectrum beyond modifying the molecular moments of inertia~\cite{grebenev1,grebenev2,HENDIreview}.

On the other hand, a low temperature superfluid medium is distinct
from the nonrelativistic vacuum in that it lacks a basic symmetry:
Galilean invariance.  The superfluid condensate defines a preferred
rest frame, an ``ether" relative to which molecular velocities should
be measured.  This implies, in particular, that the energy-momentum
relationship of molecules immersed in superfluids can deviate from a
parabolic form.  Of more interest spectroscopically, the presence of a
preferred rest frame also gives rise to the possibility of
couplings between linear and angular momenta that are strictly forbidden in an
environment with Galilean invariance.  Evidence of such couplings
would provide one of the most direct experimental signatures of the
superfluid medium.

This paper develops a simple and general theoretical framework for the
rigid-body dynamics of a molecule immersed in a bosonic superfluid. 
The classical analysis of an immersed body, presented in Section
\ref{sec1}, shows that an incompressible fluid medium adds 11
parameters to the usual 4 parameters characterizing a free rigid body. 
As a result, an immersed body is typically chaotic.  The classical
model is quantized in Section \ref{sec2}, and the quantum Hamiltonian
is discussed.  The most interesting new feature is a term of the form
${\bf p}\cdot{\bf J}$ which directly couples the linear momentum
of the molecule to its angular momentum.  This coupling requires
chirality in the molecule, and corresponds intuitively to the action
of a propeller.  It fully lifts the usual $(2J+1)$-fold degeneracy of
rigid-rotor states with angular momentum $J$, providing a clear
spectroscopic signature of the superfluid-induced
rotational-translational coupling.  An unusual consequence is that the
ground state acquires nonzero linear momentum.  In Section \ref{sec3},
the magnitude of the rotational-translational coupling is estimated
for the chiral molecules HOOH and FOOF\null.  The predicted splitting
of rotational absorption lines, at least for HOOH, is comparable to the
resolution of current experiments.  Finally, in Section
\ref{discussion}, the results are discussed in the context of related
work.

\section{Classical mechanics of an immersed rigid body}
\label{sec1}

We consider a rigid body immersed in unbounded, incompressible fluid
of uniform density $\rho_{f}$.  The body occupies a bounded, simply-connected
volume $V$ from which the fluid is excluded; the surface of this
volume is denoted by $\partial V$.  The velocity of the fluid is
described by a potential $\phi({\bf x})$, such that ${\bf v}_{f} =
-\boldgrad\phi$, which satisfies $\grad^{2}\phi = 0$ inside the fluid
volume and $|\boldgrad\phi({\bf x})| \rightarrow 0$ as $|{\bf
x}|\rightarrow\infty$.  On the moving surface $\partial V$, the fluid
velocity also must satisfy $({\bf v}_{f} - {\bf v}_{s})\cdot\nhat =
0$, where $\nhat$ is the surface normal and ${\bf v}_{s}$ is the
surface velocity.  Our goal is to obtain a classical Hamiltonian for the
immersed body which can be readily quantized.  The classical problem is
treated in many hydrodynamics texts (e.g., Ref.~\cite{lamb}),
to which we refer the reader desiring greater detail.

We first choose an origin and coordinate axes fixed in the body; this
defines the body frame.  The origin of the body frame is at ${\bf y}$
in the lab frame, and the body axes are rotated by $\hat{\cal R}$ from
the lab axes.  The three components of ${\bf y}$ and the three Euler
angles needed to parametrize the matrix $\hat{\cal R}$ completely
specify the position of the body.  When ${\bf y}={\bf 0}$ and
$\hat{\cal R} = \hat{\openone}$, the body and lab frames coincide, and
the body is said to be in the reference position.  Let $E_{t}$ be the
Euclidean transformation which moves the body from the reference
position to its position at time $t$.  Explicitly,
\begin{equation}
	E_{t}:{\bf x}\mapsto{\bf y}(t)+\hat{\cal R}(t){\bf x}.
\end{equation}
The time-dependence of $V$ is simply $V_{t} = E_{t} V_{0}$, where
$V_{0}$ is a fixed volume; similarly, $\partial V_{t} = E_{t} \partial
V_{0}$.  The lab-frame velocity of a point $E_{t}{\bf x}$, which is
fixed in the body, is
\begin{equation}
\label{ptvelocity}
	\frac{d}{dt}(E_{t}{\bf x}) = \dot{\bf y} + 
	{\bm\omega}\times(\hat{\cal R}{\bf x}),
\end{equation}
where ${\bm\omega}$ is the body's angular velocity.  The velocity
potential at a point ${\bf x}$ which is fixed in the
body frame,
$\phi'({\bf x})\equiv\phi(E_{t}{\bf x})$,
satisfies fairly simple equations: $\grad^{2}\phi' = 0$ outside
$V_{0}$, $|\boldgrad\phi'({\bf x})| \rightarrow 0$ as $|{\bf
x}|\rightarrow\infty$, and
\begin{eqnarray}
	\boldgrad\phi'({\bf x})\cdot\nhat &=& -\left(\hat{\cal 
	R}^{T}{\bf v}_{s}(E_{t}{\bf x})\right)\cdot\nhat
\nonumber\\
	&=& -\left(\hat{\cal R}^{T}\dot{\bf y} + 
	(\hat{\cal R}^{T}{\bm\omega})\times{\bf x}\right)\cdot\nhat
\end{eqnarray}
for ${\bf x}$ on the fixed surface $\partial V_{0}$, where the surface
velocity was calculated using Eq.~(\ref{ptvelocity}).  Indeed, the
equations for $\phi'$ are linear, with inhomogeneous boundary
conditions linear in $\hat{\cal R}^{T}\dot{\bf y}$ and $\hat{\cal
R}^{T}{\bm\omega}$.  The velocity potential can therefore be expressed
as a linear combination,
\begin{equation}
\label{velpot}
	\phi'({\bf x}) = 
	-(\hat{\cal R}^{T}\dot{\bf y})_{\mu}\psi_{\mu}({\bf x}) -
	(\hat{\cal R}^{T}{\bm\omega})_{\mu}\rotfnc_{\mu}({\bf x}),
\end{equation}
where the harmonic functions $\psi_{\mu}$ and $\rotfnc_{\mu}$ satisfy
\begin{subequations}
\begin{eqnarray}
\label{eq:psieqn}
	\boldgrad\psi_{\mu}({\bf x})\cdot\nhat &=& {\bf e}_{\mu}\cdot\nhat,
\\
\label{eq:rotfnceqn}
	\boldgrad\rotfnc_{\mu}({\bf x})\cdot\nhat &=&
	({\bf e}_{\mu}\times{\bf x})\cdot\nhat
\end{eqnarray}
\end{subequations}
on $\partial V_{0}$.  These six functions characterize the fluid
response to the six independent motions of the body; they depend only
on the shape of $\partial V_{0}$.

For a free rigid body, the Lagrangian is just the kinetic energy. 
When the body is immersed, this is augmented by the kinetic energy of
the fluid:
\begin{eqnarray}
\label{fluidL}
	\delta L &=& \int_{V^{c}}d^{3}x \frac{1}{2}\rho_{f}|\boldgrad\phi({\bf 
	x})|^{2}
\nonumber\\
	&=& \int_{V_{0}^{c}}d^{3}x \frac{1}{2}\rho_{f}|\boldgrad\phi'({\bf 
	x})|^{2},
\end{eqnarray}
where the second integral is over the fluid volume $V_{0}^{c}$, the
complement of the body volume $V_{0}$.  This is easy to evaluate using
Eq.~(\ref{velpot}).  The total Lagrangian, including the fluid
contribution [Eq.~(\ref{fluidL})], then has the form
\begin{equation}
\label{fullL}
	L = \frac{1}{2} 
	\begin{pmatrix}
		\dot{\bf y}^{T}\hat{\cal R} & {\bm\omega}^{T}\hat{\cal R} \cr
	\end{pmatrix}
	\begin{pmatrix}
		\hat{M} & \hat{G} \cr \hat{G}^{T} & \hat{I} \cr
	\end{pmatrix}
	\begin{pmatrix}
		\hat{\cal R}^{T}\dot{\bf y} \cr \hat{\cal R}^{T}{\bm\omega} \cr
	\end{pmatrix}
	.
\end{equation}
The mass tensor $\hat{M}$, the rotational-translational coupling
tensor $\hat{G}$, and the inertia tensor $\hat{I}$ can each be
expressed as a sum of rigid-body and fluid tensors:
$\hat{M}=\hat{M}^{(0)}+\delta\hat{M}$ and so on.

The rigid-body tensors are
\begin{subequations}
\label{eqs:rigidbodytensors}
\begin{eqnarray}
	M^{(0)}_{\mu\nu} &=& \int d^{3}x \rho({\bf x}) \delta_{\mu\nu},
\\
	G^{(0)}_{\mu\nu} &=& \int d^{3}x \rho({\bf x})\epsilon_{\mu\nu\lambda}
	x_{\lambda},
\\
	I^{(0)}_{\mu\nu} &=& \int d^{3}x \rho({\bf x})\left(|{\bf x}|^{2}
	\delta_{\mu\nu} - x_{\mu}x_{\nu}\right),
\end{eqnarray}
\end{subequations}
where $\rho({\bf x})$ is the body-frame mass density of the body.  The
expressions for $\hat{M}^{(0)}$ and $\hat{I}^{(0)}$ are the usual
ones, and $\hat{G}^{(0)}$ is typically made to vanish by choosing the
body's center of mass as the origin of the body frame.  The rigid-body
Lagrangian then takes the familiar form
\begin{equation}
	L^{(0)} = 
	\frac{1}{2}\dot{\bf y}^{T}\hat{M}^{(0)}\dot{\bf y} + 
	\frac{1}{2}{\bm\omega}^{T}\hat{\cal R}\hat{I}^{(0)}\hat{\cal 
	R}^{T}{\bm\omega}.
\end{equation}
The fluid tensors are
\begin{subequations}
\label{eqs:fluidtensors}
\begin{eqnarray}
	\delta M_{\mu\nu} &=& \int_{V_{0}^{c}}d^{3}x 
	\rho_{f}\boldgrad\psi_{\mu}({\bf x}) \cdot
	\boldgrad\psi_{\nu}({\bf x}),
\\
	\delta G_{\mu\nu} &=& \int_{V_{0}^{c}}d^{3}x
	\rho_{f}\boldgrad\psi_{\mu}({\bf x}) \cdot
	\boldgrad\rotfnc_{\nu}({\bf x}),
\\
	\delta I_{\mu\nu} &=& \int_{V_{0}^{c}}d^{3}x
	\rho_{f}\boldgrad\rotfnc_{\mu}({\bf x}) \cdot
	\boldgrad\rotfnc_{\nu}({\bf x}).
\end{eqnarray}
\end{subequations}
Note that the mass and inertia tensors $\hat{M}$ and $\hat{I}$ are
symmetric, while the rotational-translational coupling $\hat{G}$ has
no obvious symmetry.  In the most general case there are $6+6+9=21$
parameters in the Lagrangian.  We can place the body origin at a
conveniently chosen point in the body, reducing this to 18.  We can
then rotate the body axes with respect to the body, leaving 15.  By
comparison, the free rigid rotor has 4 parameters.

The classical Hamiltonian follows from the Lagrangian
[Eq.~(\ref{fullL})]:
\begin{equation}
\label{fullH}
	{\cal H} = \frac{1}{2}
	\begin{pmatrix}
		{\bf p}^{T}\hat{\cal R} & {\bf J}^{T}\hat{\cal R} \cr
	\end{pmatrix}
	\begin{pmatrix}
		\hat{M} & \hat{G} \cr \hat{G}^{T} & \hat{I} \cr
	\end{pmatrix}
	^{-1}
	\begin{pmatrix}
		\hat{\cal R}^{T}{\bf p} \cr \hat{\cal R}^{T}{\bf J} \cr
	\end{pmatrix}
	,
\end{equation}
where ${\bf p}$ is the linear momentum, canonically conjugate to ${\bf
y}$, and ${\bf J}$ is the angular momentum in the lab frame. 
The components of ${\bf J}$ have Poisson brackets $\left[ J_{\mu},
J_{\nu} \right]_{\rm cl} = \epsilon_{\mu\nu\lambda}J_{\lambda}$.  It
is convenient also to work with ${\bf J}' = \hat{\cal R}^{T}{\bf J}$
and ${\bf p}' = \hat{\cal R}^{T}{\bf p}$, the angular and linear
momenta in the body frame, in terms of which the Hamiltonian is
particularly simple.  These variables have Poisson brackets $\left[
J'_{\mu}, J'_{\nu} \right]_{\rm cl} =
-\epsilon_{\mu\nu\lambda}J'_{\lambda}$ and $\left[ J'_{\mu}, p'_{\nu}
\right]_{\rm cl} = -\epsilon_{\mu\nu\lambda}p'_{\lambda}$.  The
classical Poisson brackets relate to the quantum commutators in the
usual way, as $[A,B] = i[A,B]_{\rm cl}$.

To express Eq.~(\ref{fullH}) in a form more amenable to quantization,
we first define new tensors $\hat\alpha$, $\hat\beta$, and $\hat\gamma$
through
\begin{equation}
	\begin{pmatrix}
		\hat{\alpha} & \hat{\beta} \cr \hat{\beta}^{T} & \hat{\gamma} \cr 
	\end{pmatrix}
	\equiv
	\begin{pmatrix}
		\hat{M} & \hat{G} \cr \hat{G}^{T} & \hat{I} \cr 
	\end{pmatrix}
	^{-1}.
\end{equation}
In the following section it is shown that the model can be
quantized without complication provided that $\hat\beta$ is
symmetric. 
We can ensure that $\hat\beta$ is in fact symmetric by choosing the
body frame's origin correctly.  For a free rigid body, the body's
center of mass is the correct choice, but this is not true in general.
The tensors $\hat\alpha$ and
$\hat\beta$ transform in a particular way when the
body origin is translated within the body ($\hat\gamma$ is unchanged);
we can use the transformation law to find the appropriate choice of origin.  
Specifically, consider shifting the body
origin by ${\bf d}$, so that all body-frame coordinates transform
as ${\bf x}\mapsto{\bf x}-{\bf d}$.  Then a direct calculation from
Eqs.~(\ref{eqs:rigidbodytensors}) shows that the rigid-body tensors
transform as
\begin{equation}
	\begin{pmatrix}
		\hat{M}^{(0)} & \hat{G}^{(0)} \cr \hat{G}^{(0)T} & \hat{I}^{(0)} \cr
	\end{pmatrix}
	\mapsto
	\begin{pmatrix}
		\hat{\openone} & 0 \cr \epsilon{\bf d} & \hat{\openone} \cr
	\end{pmatrix}
	\begin{pmatrix}
		\hat{M}^{(0)} & \hat{G}^{(0)} \cr \hat{G}^{(0)T} & \hat{I}^{(0)} \cr
	\end{pmatrix}
	\begin{pmatrix}
		\hat{\openone} & -\epsilon{\bf d} \cr 0 & \hat{\openone} \cr
	\end{pmatrix},
\end{equation}
where the antisymmetric matrix $\epsilon{\bf d}$ has matrix elements $(\epsilon{\bf d})_{\mu\nu} = \epsilon_{\mu\nu\lambda}d_{\lambda}$.  It is not hard to see that the fluid tensors
transform in exactly the same way.  Inverting the relation, we find that $\hat\beta \mapsto \hat\beta + (\epsilon{\bf d})\hat\gamma$.  Now, the vector ${\bf b}$ with components $b_{\lambda} = \epsilon_{\mu\nu\lambda}\beta_{\mu\nu}$ is the zero vector exactly when $\hat\beta$ is symmetric, and it transforms as
${\bf b} \mapsto {\bf b} + \left\{({\rm Tr}{\hat\gamma})\hat{\openone} - \hat\gamma\right\}{\bf d}$.
The matrix multiplying ${\bf d}$ is positive definite, because $\hat\gamma$ is, so it can be inverted.
The result is that if $\hat\beta$ is not symmetric, it can be made so by
shifting the origin of the body frame by ${\bf d}
= -\left\{({\rm Tr}{\hat\gamma})\hat{\openone} - \hat\gamma\right\}^{-1}{\bf b}$ within the body.
We will assume that this has been done.

With the body origin fixed, rotation of the body axes can be used
to diagonalize $\hat{\gamma}$, leaving it in the form $\hat\gamma =
{\rm diag}(2A,2B,2C)$.  (Since $\hat\beta$ transforms as a tensor
under rotation, it will remain symmetric.)
Note that $A$, $B$, and $C$ are \emph{effective} rotational constants,
shifted from their rigid-body values by the body's interaction
with the fluid.
Finally, the symmetric matrices $\hat\alpha$
and $\hat\beta$ can be broken down into their scalar and rank-2
spherical tensor components:
\begin{equation}
	\hat\alpha = \alpha^{(0)}_{0}\hat{\openone} + 
	\sum_{q=-2}^{2}\alpha^{(2)}_{q}\hat{\cal M}^{(2)}_{q},
\end{equation}
and analogously for $\hat\beta$, where
\begin{equation}
	\hat{\cal M}^{(2)}_{0} = \frac{1}{\sqrt{6}}
	\begin{pmatrix}
		-1 &    &    \cr 
		   & -1 &    \cr
		   &    & 2  \cr
	\end{pmatrix}
	\! ,\;
	\hat{\cal M}^{(2)}_{\pm 1} = -\frac{1}{2}
	\begin{pmatrix}
		0 & 0 & \pm 1 \cr 
		0 & 0 & i \cr
		\pm 1 & i & 0 \cr
	\end{pmatrix}
	\! , \nonumber\\
\end{equation}
\begin{equation}
	\hat{\cal M}^{(2)}_{\pm 2} = \frac{1}{2}
	\begin{pmatrix}
		1 & \pm i & 0 \cr
		\pm i & -1 & 0 \cr
		0 & 0 & 0 \cr
	\end{pmatrix}
	\! .
\end{equation}
These matrices are defined to have nice rotational properties; in
particular,
\begin{equation}
	\hat{\cal R}\hat{\cal M}^{(2)}_{q}\hat{\cal R}^{T} = 
	\sum_{p}\hat{\cal M}^{(2)}_{p}{\cal D}^{(2)}_{pq},
\end{equation}
where the ${\cal D}^{(2)}_{pq}$ are rotation matrices, given by known
functions of the Euler angles~\cite{brink}. The Hamiltonian then has
the form
\begin{eqnarray}
\label{ham1}
	{\cal H} &=&
	\frac{1}{2}\alpha^{(0)}_{0}p^{2} +
	\beta^{(0)}_{0}{\bf p}\cdot{\bf J} +
	{\cal H}_{rr}(A,B,C) \nonumber\\ &&\;\;\;+
	\sum_{q} \frac{1}{2}\alpha^{(2)}_{q}\Pi^{(2)}_{q} +
	\sum_{q} \beta^{(2)}_{q}\Theta^{(2)}_{q},
\end{eqnarray}
where ${\cal H}_{rr}$ is the rigid rotor Hamiltonian,
\begin{equation}
	{\cal H}_{rr}(A,B,C) = A {J_{x}'}^{2} + B {J_{y}'}^{2} + C {J_{z}'}^{2},
\end{equation}
and $\Pi^{(2)}$ and $\Theta^{(2)}$ are spherical tensors of rank 2,
with components
\begin{equation}
\label{tensorops1}
	\Pi^{(2)}_{q} = ({\bf p}')^{T}\hat{\cal 
	M}^{(2)}_{q}{\bf p}' ; \qquad \Theta^{(2)}_{q} = ({\bf 
	p}')^{T}\hat{\cal M}^{(2)}_{q}{\bf J}'.
\end{equation}
More precisely, $\Pi^{(2)}$ and $\Theta^{(2)}$ transform as spherical
tensors of rank 2 under rotations generated by $-{\bf J}'$.

Before proceeding to the quantum case, we consider the classical
dynamics of the model.  The Lagrangian is invariant under arbitrary
translations (${\bf y} \mapsto {\bf y}+{\bf a}$) and system rotations
(${\bf y} \mapsto \hat{\cal O}{\bf y}$, $\hat{\cal R} \mapsto
\hat{\cal O}\hat{\cal R}$), but not under body rotations ($\hat{\cal
R} \mapsto \hat{\cal O}\hat{\cal R}$) alone.  Therefore ${\bf p}$ and
the total (spin plus orbital) angular momentum ${\bf J} + {\bf
y}\times{\bf p}$ are constants of the motion, but ${\bf J}$ is not, in
contrast to the free rigid rotor.  We can use these symmetries to
simplify the system.  Four constants of the motion, which have
vanishing Poisson brackets with ${\cal H}$ and with each other, are
${\bf p}\cdot{\bf J}$ and the components of ${\bf p}$.  By fixing
these constants and eliminating the coordinates which are conjugate to
them (the components of ${\bf y}$ and one of the three Euler angles),
we restrict the Hamiltonian to a reduced phase space; the equations of
motion for the remaining degrees of freedom are unchanged by the
reduction.  In this case, the original six degrees of freedom are
reduced to two, so the reduced phase space is four-dimensional.  It
can be parametrized by the directions of ${\bf p}'$ and ${\bf J}'$,
expressed in polar coordinates by the four angles
$\left(\theta_{p},\phi_{p},\theta_{J},\phi_{J}\right)$.
The energy shells of the reduced phase space (i.e., the level sets
of ${\cal H}$) are three-dimensional.

\begin{widetext}

\begin{figure}
\begin{center}
\includegraphics[width=5.5 in]{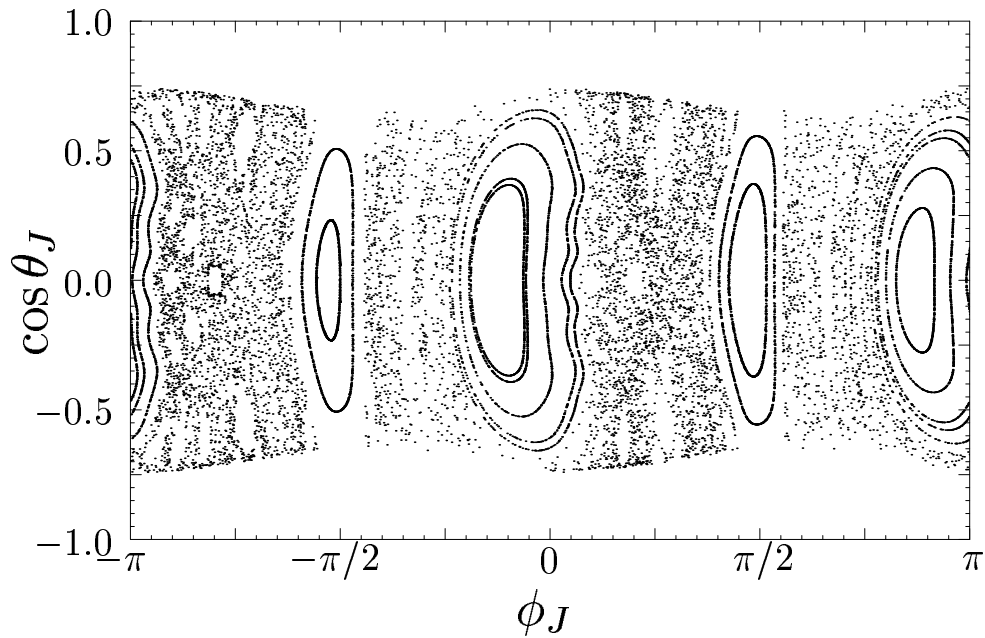}
\end{center}
\caption{\label{fig:poincare}Poincar\'e section of the reduced phase
space of an immersed body, showing both chaotic and quasi-integrable
regions.}
\end{figure}

\end{widetext}

The integrability of a Hamiltonian system with two degrees of freedom
can be ascertained using a Poincar\'e section, essentially by
inspection.  To illustrate that the immersed rigid body is generally
chaotic, we display a Poincar\'e section of the reduced phase space in
Fig.~\ref{fig:poincare}.  The figure was generated by repeatedly
integrating the equations of motion, starting at many different points
within the ${\cal H}=2$ energy shell, and plotting the point
$\left(\phi_{J},\cos\theta_{J}\right)$ each time a trajectory
crossed the plane $\theta_{p}=\pi/2$.  The Hamiltonian parameters were
$\hat\alpha = {\hat{\openone}}$, $\hat\beta = \text{diag}(1,1,-2)$, and
$\hat\gamma = \text{diag}(1,2,\sqrt{5})$; the conserved quantities
were set to ${\bf p}\cdot{\bf J} = 0.1$ and ${\bf p} = 0.1{\bf
e}_{z}$.  If the system were integrable,
each phase space trajectory would be
confined to a two-dimensional torus, and would appear in the figure as
a densely dotted curve.  Instead, the figure shows that the phase
space contains chaotic regions, where trajectories ergodically visit
three-dimensional volumes, as well as quasi-integrable regions, where
trajectories are confined to two-dimensional tori.  It should be noted that this
set of parameter values corresponds to very strong coupling
between linear and angular momenta, and was chosen to highlight the
presence of chaos.  The physical cases we will consider
in this paper are much more nearly integrable.  Nevertheless, the existence
of classical chaos suggests that the quantum spectrum, unlike that of the
free rigid rotor, cannot be determined algebraically.  We will need to use an
approximate method, in this case a perturbative expansion in powers of
the linear momentum.

\section{Quantum Hamiltonian}
\label{sec2}

The quantization of the classical Hamiltonian [Eq.~(\ref{ham1})] is
straightforward, and proceeds as in any standard angular momentum text
(see, for instance, Ref.~\cite{brink}).  There is only one
subtlety.  Since ${\bf p}'$ and ${\bf J}'$ have a nonvanishing Poisson
bracket, their quantum counterparts do not commute, and the
quantization of terms like $p'_{\mu}J'_{\nu}$ suffers from an ordering
ambiguity.  We can sidestep this problem because we only need to
quantize the components of ${\Theta}^{(2)}$: since $\left[{J}'_{\mu},
{p}'_{\nu}\right]_{\rm cl}$ is $\mu\nu$-antisymmetric, while each
matrix $\hat{\cal M}^{(2)}_{q}$ is symmetric, the operators
${\Theta}^{(2)}_{q}$ are independent of the relative ordering of ${\bf
p}'$ and ${\bf J}'$.  For the remainder of the paper, ${\bf p}$,
${\cal H}$, etc., will refer to the appropriate quantum operators (or
their eigenvalues) unless otherwise noted.

Breaking the Hamiltonian down into ${\cal H}_{0}$ and $\delta{\cal
H}$, respectively the first three and last two terms of
Eq.~(\ref{ham1}), allows us to proceed perturbatively.  Since
$\delta{\cal H}$ vanishes for a free rigid rotor, this separates out
the effects of the fluid.  The ${\bf p}\cdot{\bf J}$ term also is
absent for a free rotor, but this term commutes with $p^{2}$ and
${\cal H}_{rr}$, so it can be treated exactly.

Because ${\bf p}$ is conserved, we can fix ${\bf p} = k{\bf e}_{z}$,
$k \ge 0$, with no loss of generality.  A convenient basis for the
remaining (angular) degrees of freedom is the set of simultaneous
eigenvectors of ${J}^{2}$, ${J}'_{z}$, and ${J}_{z}$; the eigenvector
with respective eigenvalues $J(J+1)$, $K$, and $M$ is denoted by
$\left|JKM\right>$.  The tensor operators [Eq.~(\ref{tensorops1})]
become
\begin{subequations}
\begin{eqnarray}
	\!\!\!\!\!\!\!\!\! {\Pi}^{(2)}_{q} \!\! &=& \!\! \frac{2}{\sqrt{6}}k^{2}{\cal D}^{(2)}_{0q},
\\
	\!\!\!\!\!\!\!\!\! {\Theta}^{(2)}_{q} \!\! &=& \!\! k\left( \frac{1}{2}{J}_{-}{\cal 
	D}^{(2)}_{-1q} + \frac{2}{\sqrt{6}}{J}_{z}{\cal D}^{(2)}_{0q} - 
	\frac{1}{2}{J}_{+}{\cal D}^{(2)}_{1q} \right),
\end{eqnarray}
\end{subequations}
where $J_{\pm}\equiv J_{x}\pm iJ_{y}$.  Their matrix elements are
expressible in terms of Wigner 3-$j$ symbols~\cite{brink}:
\begin{widetext}
\begin{subequations}
\begin{eqnarray}
\left<J_{2}K_{2}M_{2}\right|\left.{\Pi}^{(2)}_{q}\right.\left|J_{1}K_{1}M_{1}\right>
&=&
\frac{2}{\sqrt{6}}k^{2}(-1)^{K_{1}+M_{1}}\sqrt{(2J_{2}+1)(2J_{1}+1)} 
\begin{pmatrix}
	J_{2} & 2 & J_{1} \cr K_{2} & q & -K_{1} \cr
\end{pmatrix}
\begin{pmatrix}
	J_{2} & 2 & J_{1} \cr M_{2} & 0 & -M_{1} \cr
\end{pmatrix}
\nonumber\\
&=&
(-1)^{K_{1}}
\begin{pmatrix}
	J_{2} & 2 & J_{1} \cr K_{2} & q & -K_{1} \cr
\end{pmatrix}
\left< {J_{2}M_{2}\left| {\left|{\Pi}^{(2)}\right|} \right|J_{1}M_{1}} \right>, 
\end{eqnarray}
and
\begin{eqnarray}
\left<J_{2}K_{2}M_{2}\right|\left.{\Theta}^{(2)}_{q}\right.\left|J_{1}K_{1}M_{1}\right>
&=&
k(-1)^{K_{1}+M_{1}}\sqrt{(2J_{2}+1)(2J_{1}+1)}
\begin{pmatrix}
	J_{2} & 2 & J_{1} \cr K_{2} & q & -K_{1} \cr
\end{pmatrix}
\times \nonumber\\ 
&& \Bigg\{
\frac{1}{2}\sqrt{(J_{2}-M_{2})(J_{2}+M_{2}+1)}
\begin{pmatrix}
	J_{2} & 2 & J_{1} \cr M_{2}+1 & -1 & -M_{1} \cr
\end{pmatrix}
+
\frac{2 M_{2}}{\sqrt{6}}
\begin{pmatrix}
	J_{2} & 2 & J_{1} \cr M_{2} & 0 & -M_{1} \cr
\end{pmatrix} 
\nonumber\\
&&\;\;\;\;-\frac{1}{2}\sqrt{(J_{2}-M_{2}+1)(J_{2}+M_{2})}
\begin{pmatrix}
	J_{2} & 2 & J_{1} \cr M_{2}-1 & 1 & -M_{1} \cr
\end{pmatrix}
\Bigg\} \nonumber\\
&=& (-1)^{K_{1}}
\begin{pmatrix}
	J_{2} & 2 & J_{1} \cr K_{2} & q & -K_{1} \cr
\end{pmatrix}
\left< {J_{2}M_{2}\left| {\left|{\Theta}^{(2)}\right|} 
\right|J_{1}M_{1}} \right>.
\end{eqnarray}
\end{subequations}
\end{widetext}
The simple dependence of the matrix elements on $K_{1}$, $K_{2}$ and
$q$ is a consequence of the tensorial nature of the operators, and is
guaranteed by the Wigner-Eckart theorem.  The reduced matrix elements
vanish unless $M_{2}=M_{1}$, by inspection, so $J_{z}$ commutes with
both ${\cal H}_{0}$ and $\delta{\cal H}$.  This is to be expected from
the classical conservation of ${\bf p}\cdot{\bf J}$, which equals
$kJ_{z}$ for our choice of ${\bf p}$.

The eigenvectors of ${\cal H}_{0}$ are the simultaneous eigenvectors
of the rigid rotor Hamiltonian and ${J}_{z}$.  Since ${\cal H}_{rr}$
can be written in a normalized form,
\begin{equation}
	{\cal H}_{rr}(A,B,C) = \frac{1}{2}(A+C){J}^{2} + 
	\frac{1}{2}(A-C){\cal H}_{rr}(1,\kappa,-1),
\end{equation}
the eigenvectors depend only on the asymmetry parameter $\kappa =
(-A+2B-C)/(A-C)$.  The eigenstates evolve continuously with $\kappa$;
they become eigenstates of ${J}'_{x}$ and ${J}'_{z}$ in the prolate
($\kappa\rightarrow -1$) and oblate ($\kappa\rightarrow 1$) limits
respectively, with corresponding eigenvalues $K_{-1}$ and $K_{+1}$. 
Following standard notation, we denote these states as $\left|\tau
M\right>$, where $\tau \in \left\{ J_{K_{-1}K_{+1}} \right\} = \left\{
0_{00}, 1_{01}, 1_{11}, 1_{10}, \ldots \right\}$ indicates both the
angular momentum and the two limiting values of $K$, and $M$ is the
eigenvalue of $J_{z}$.  The unperturbed energies are
\begin{eqnarray}
	E^{(0)}_{\tau M} &=& \frac{1}{2}\alpha^{(0)}_{0}k^{2} + \beta^{(0)}_{0} M 
	k + \frac{1}{2}(A+C)J(J+1) \nonumber\\ &&\;\;\;+
	\frac{1}{2}(A-C){\cal E}_{\tau}(\kappa),
\end{eqnarray}
written in terms of the normalized rigid rotor eigenvalues ${\cal
E}_{\tau}(\kappa)$.

We are now in a position to find the perturbed energies to any order
in $\delta{\cal H}$.  The expansion can be reordered as an expansion
in the momentum $k$.  The first-order correction comes from the
diagonal elements of $\Theta^{(2)}$, which turn out to be proportional
to $M$ for each $\tau$.  A given level has energy
\begin{eqnarray}
\label{levelenergy}
E_{\tau M}(k) &=& 
\frac{1}{2}(A+C)J(J+1) +
\frac{1}{2}(A-C){\cal E}_{\tau}(\kappa)
\nonumber\\ && + \tilde{B}_{\tau} M k + \frac{1}{2}\tilde{A}_{\tau M}k^{2} + 
O(k^{3}),
\end{eqnarray}
where $\tilde{B}_{\tau}$ and $\tilde{A}_{\tau M}$ are 
calculated from matrix elements of $\delta{\cal{H}}$ in the usual
way.  In particular,
\begin{subequations}
\begin{eqnarray}
\tilde{B}_{\tau} &=& \beta^{(0)}_{0} + \frac{1}{Mk}\sum_{q}\beta^{(2)}_{q}\left(\Theta^{(2)}_{q}\right)_{\tau\tau M}, \\
\tilde{A}_{\tau M} &=& \alpha^{(0)}_{0} + \frac{1}{k^{2}}\sum_{q}\alpha^{(2)}_{q}\left(\Pi^{(2)}_{q}\right)_{\tau\tau M}
\nonumber\\ && +
\frac{2}{k^{2}}\sum_{q,q'}\beta_{q'}^{(2)}\beta_{q}^{(2)}\sum_{\tau'}\frac{\big(\Theta^{(2)}_{q'}\big)_{\tau\tau' M}
\big(\Theta^{(2)}_{q}\big)_{\tau'\tau M}}{E^{(0)}_{\tau M} - E^{(0)}_{\tau' M}}, \nonumber\\
\end{eqnarray}
\end{subequations}
where $(\Theta)_{\tau\tau' M}$ is shorthand for $\left<\tau M\left|\Theta\right|\tau' M\right>$.
Higher-order terms are suppressed by additional factors
of $k{\beta}^{(2)}_{q}$ or $k^{2}{\alpha}^{(2)}_{q}$; for the cases we
consider in the next section, these factors are less than $10^{-3}$,
so it is reasonable to truncate the expansion here.  Clearly
$\tilde{A}_{\tau M}^{-1}$ is an effective mass; it reduces to the mass
of the body in the absence of the fluid.  The constant
$\tilde{B}_{\tau}$ is a pseudoscalar associated with the family of
levels $\left|\tau M\right>$.  It is nonzero only in the presence of
the fluid, and only when the immersed body is \emph{chiral}.  Judging
from the spectrum, $\tilde{B}_{\tau}$ measures the tendency of the
immersed body to have its linear and angular momenta aligned.  When
this tendency is strong, the body behaves as a kind of ``quantum
propeller.''

\section{Molecular surface model and numerical results}
\label{sec3}

In order to apply these results to a real system, we must specify the
fluid density $\rho_{f}$, model the surface $\partial V_{0}$ and mass
density $\rho({\bf x})$ which characterize the immersed body, and
calculate the tensors $\hat{M}$, $\hat{G}$, and $\hat{I}$ from these
inputs.  In this section we consider the interesting case of an
immersed \emph{molecule} in superfluid helium.  The fluid density
$\rho_{f}$ is approximately $0.1$\;amu-\AA$^{-3}$.  The mass density of the
molecule can be represented as $\rho({\bf x}) =
\sum_{i}m_{i}\delta({\bf x}-{\bf x}_{i})$, where the atomic
coordinates ${\bf x}_{i}$ are known.  Helium is kept away from the
molecule primarily by the short-ranged Fermi repulsion between
molecular electrons and helium electrons.  This results in a smoothly
varying helium density, interpolating from zero near the molecule to
$\rho_{f}$ far away.  Our model approximates this smooth variation by
a discontinuous jump, localized on the imaginary surface $\partial
V_{0}$.  The optimal choice of surface is not obvious, but some
criteria are clearly important.  The surface should share the
symmetries of the molecule; it should be smooth, since the electronic
densities are smooth; and its size and shape should be physically
reasonable: when compared to the actual helium density profile, the
surface should approximate the surface $\rho = \frac{1}{2}\rho_{f}$. 
With these criteria in mind, we model the surface $\partial V_{0}$ for
an immersed molecule in the following way.  First, we place a sphere
of radius $R_{i}$ at each atomic coordinate ${\bf x}_{i}$.  Each
radius is proportional to the van der Waals distance between a helium
atom and an atom of type $i$: $R_{i} = c R_{\text{He-}i}$, where the
parameter $c$ will be of order $1$.  (For our calculations, we used
$R_{\text{He-H}}=2.60\text{\;\AA}$,
$R_{\text{He-F}}=2.87\text{\;\AA}$, and
$R_{\text{He-O}}=2.92\text{\;\AA}$~\cite{vanderwaalsref}.) The union of
these spheres forms a cuspy volume.  The surface radius of the body
volume, relative to an origin within the molecule, is represented in
polar coordinates as $r_{s}(\theta,\phi)$.  We then smooth the
function $r_{s}$ using the rotationally invariant linear operator
which maps $Y_{lm}(\theta,\phi)\mapsto \exp\left(-\alpha
l\right)Y_{lm}(\theta,\phi)$, with $\alpha=0.1$, to yield the final
surface $\partial V_{0}$.  (Many other smoothing operators would do
as well; our results are not very sensitive to the choice.)  This
prescription gives a single-parameter
family of smooth surfaces which share the symmetry of the molecule;
the parameter $c$ can be adjusted to exclude superfluid from an
appropriately sized volume.

Once the surface is fixed, the fluid tensors can be calculated either
numerically or analytically.  We used a perturbative analytical
method, valid for nearly spherical surfaces, where the surface radius
is a weakly varying function of polar angle: $r_{s}(\theta,\phi) =
r_{0}\left(1+\epsilon(\theta,\phi)\right)$, with $\epsilon \ll 1$. 
The fluid response functions $\psi_{\mu}$ and $\rotfnc_{\mu}$ (see
Section \ref{sec1}) can then be calculated as power series in
$\epsilon$, as can the fluid contributions to the tensors.  This
calculation has been carried through to second order, which is the
lowest order at which nontrivial rotational-translational coupling
(i.e. a nonzero value for $\tilde{B}_{\tau}$) is seen.  The
calculation and results are described in the Appendix.

\begin{figure}
\begin{center}
\includegraphics[width=3.25 in]{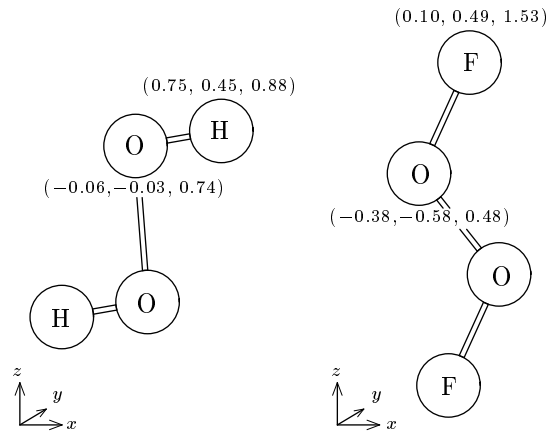}
\end{center}
\caption{\label{fig:mols}
Structures of the chiral molecules HOOH and
FOOF\null.  Atomic coordinates $(x,y,z)$ are shown, in \AA, for
two atoms
in each molecule.  The
undisplayed coordinates
are obtained from these
by a symmetry transformation, $R_{y}(\pi):(x,y,z)\mapsto(-x,y,-z)$.}
\end{figure}

We present data for the low-lying levels of two different molecules:
hydrogen peroxide (HOOH) and dioxygen difluoride (FOOF).  These
molecules are depicted in Fig.~\ref{fig:mols}, with structural
parameters taken from Ref.~\cite{molstructures}.  Each has a
$C_{2}$ point group, which we have taken to be $C_{2}(y)$ by an
appropriate rotation of body axes.  The allowed rotational levels all
have the same parity under rotation by $\pi$ around the $y$-axis:
either even ($0_{00}, 1_{11}, 2_{02},\ldots$) or odd ($1_{01}, 1_{10},
2_{12},\ldots$), depending on the symmetry of the joint
electronic/nuclear wavefunction.  In the case where allowed levels are
odd, the $J=0$ state is forbidden, so the ground state of the immersed
molecule has $J=1$, and, by Eq.~(\ref{levelenergy}), a nonzero linear
momentum.  The ground state momentum is $k_{0} = |\tilde{B}_{\tau_{0}}
/ \tilde{A}_{\tau_{0},+1}|$, where $\tau_{0}$ is the ground state with
$J=1$.

The results for HOOH and FOOF are presented in Tables~\ref{tab:h2o2}
and \ref{tab:f2o2} respectively.
(Energies are given in cm$^{-1}$ or K; these conversions are made using
$k_{B} = hc = 1$, consistent with usual spectroscopic notation.
Momenta are given in \AA$^{-1}$; this conversion is made using $\hbar = 1$.)
The surfaces considered can be parametrized by their mean radius
$r_{0}$, shown in the first column.  The values displayed correspond
to the parameter range $0.5\le c \le 1.5$ for each molecule.  Although
the inverse effective masses $\tilde{A}_{\tau M}$ are level-dependent,
this dependence is very weak, and $\tilde{A}_{\tau M} \approx
\alpha_{0}^{(0)}$ for the tabulated cases.  This approximate value is
displayed, as $\tilde{A}$, in the second column.  The remaining
columns list the zero-momentum energies $E_{\tau}(0)$ and chiral
splitting constants $\tilde{B}_{\tau}$ for the $C_{2}(y)$-odd levels
$\tau=1_{01}$ and $\tau=1_{10}$.

\begin{table}
\caption{\label{tab:h2o2}Results for HOOH\null.}
\begin{center}
\begin{tabular}{ccccccc}
$r_{0}$  &
$\tilde{A}$  &
$E_{1_{01}}(0)$  &
$\tilde{B}_{1_{01}}$ &
$E_{1_{10}}(0)$  &
$\tilde{B}_{1_{10}}$
\\
\hline
1.86  &  0.94  &  1.64  &  -2$\times 10^{-4}$  &  9.68  &  3.2$\times 10^{-3}$  \\
2.47  &  0.89  &  1.61  &  -2$\times 10^{-4}$  &  8.94  &  4.0$\times 10^{-3}$ \\
3.06  &  0.83  &  1.57  &  -3$\times 10^{-4}$  &  8.28  &  6.2$\times 10^{-3}$  \\
3.63  &  0.76  &  1.52  &  -5$\times 10^{-4}$  &  7.73  &  8.1$\times 10^{-3}$  \\
4.21  &  0.67  &  1.46  &  -6$\times 10^{-4}$  &  7.35  &  9.1$\times 10^{-3}$  \\
4.78  &  0.59  &  1.38  &  -5$\times 10^{-4}$  &  7.13  &  8.6$\times 10^{-3}$  \\
\hline
$(\text{\AA})$ & $(\text{cm}^{-1}\text{-\AA}^{2})$ & $(\text{cm}^{-1})$ & $(\text{cm}^{-1}\text{-\AA})$ &
$(\text{cm}^{-1})$ & $(\text{cm}^{-1}\text{-\AA})$
\end{tabular}
\end{center}
\end{table}

\begin{table}
\caption{\label{tab:f2o2}Results for FOOF\null.}
\begin{center}
\begin{tabular}{ccccccc}
$r_{0}$  &
$\tilde{A}$  &
$E_{1_{01}}(0)$  &
$\tilde{B}_{1_{01}}$ &
$E_{1_{10}}(0)$  &
$\tilde{B}_{1_{10}}$
\\
\hline
1.97  &  0.46  &  0.31  &  3$\times 10^{-4}$  &  0.83  &  -5$\times 10^{-4}$  \\  
2.68  &  0.42  &  0.29  &  2$\times 10^{-4}$  &  0.36  &  -2$\times 10^{-4}$  \\
3.32  &  0.43  &  0.30  &  2$\times 10^{-4}$  &  0.81  &  -6$\times 10^{-4}$  \\
3.93  &  0.40  &  0.29  &  2$\times 10^{-4}$  &  0.80  &  -6$\times 10^{-4}$  \\
4.53  &  0.37  &  0.28  &  2$\times 10^{-4}$  &  0.78  &  -6$\times 10^{-4}$  \\
5.13  &  0.34  &  0.27  &  2$\times 10^{-4}$  &  0.77  &  -6$\times 10^{-4}$  \\
\hline
$(\text{\AA})$ & $(\text{cm}^{-1}\text{-\AA}^{2})$ & $(\text{cm}^{-1})$ & $(\text{cm}^{-1}\text{-\AA})$ &
$(\text{cm}^{-1})$ & $(\text{cm}^{-1}\text{-\AA})$
\end{tabular}
\end{center}
\end{table}

To put the results in context, consider the physical parameters of
current nano\-droplet experiments, together with the hydrogen peroxide
results.  Ignoring finite-size effects, the thermal momentum
distribution is expected to be Maxwellian, i.e. $dP/dk =
Ck^{2}\exp\left(-\frac{1}{2}\tilde{A} k^{2} / T\right)$.  The
nanodroplet temperature is $T = 0.4$\;K.
This corresponds to a typical thermal momentum of $k_{rms} =
1$\;\AA$^{-1}$ for HOOH\null.  The ground-state momentum $k_{0}$, on
the other hand, is only about $0.01$\;\AA$^{-1}$, and is completely
overwhelmed by thermal fluctuations.  The $\tau=1_{01}$ and
$\tau=1_{10}$ energy levels will each acquire fine structure by
splitting into three evenly spaced sublevels with different values of
$M$; we estimate the spacing for these two sets of sublevels to be
$2$-$5\times10^{-4}$\;cm$^{-1}$ and $3$-$8\times10^{-3}$\;cm$^{-1}$ respectively.  Moreover, the
center level ($M=0$) in each triplet is sharp, while the wings ($M=\pm
1$) are broadened by the thermal spread in momentum.  Because the
experimental linewidth for rovibrational transitions is small, on the
order of $0.01$\;cm$^{-1}$, we expect this fine structure to be nearly resolvable
in the $1_{01}\rightarrow 1_{10}$ absorption line for HOOH\null.  If resolved, the
line would appear as a sharp central peak, with broader peaks
symmetrically placed on either side, at $\pm 0.003-0.008$\;cm$^{-1}$.

\begin{widetext}

\begin{figure}
\begin{xy}
	\WARMprocessMMA{fig3}{eps}{mbb}
	\xyMarkedImport{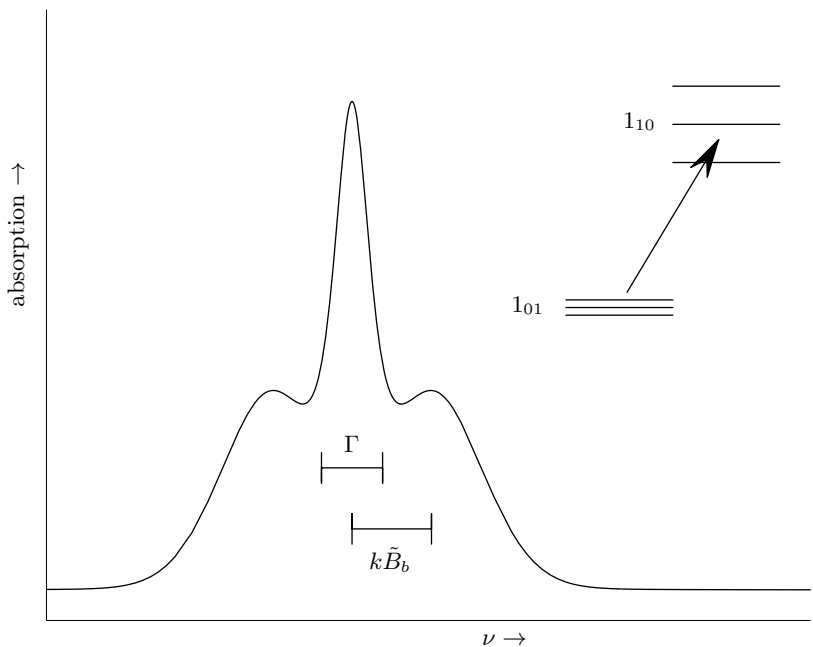}

	\xyMarkedPos{absorption}*+!R(1.5)!D(1.0)\txt{\rotatebox{90}{absorption $\rightarrow$}}
	\xyMarkedPos{absorption}*+!R(13.0)!D(1.0)\txt{\;}
	\xyMarkedPos{nu}*+!U(1.5)\txt{$\nu\rightarrow$}
	\xyMarkedPos{init}*+!R(0.3)\txt{$1_{01}$}
	\xyMarkedPos{final}*+!R(0.3)\txt{$1_{10}$}
	\xyMarkedPos{gamma}*+!D(1.5)\txt{$\Gamma$}
	\xyMarkedPos{b}*+!U(1.5)\txt{$k \tilde{B}_{b}$}
\end{xy}
\caption{\label{fig:fig43}Schematic representation of the predicted
$1_{01}\rightarrow 1_{10}$ absorption line in hydrogen peroxide.
The strength of absorption is plotted as a function of the photon
wavenumber, $\nu$.}
\end{figure}

\end{widetext}

More generally, the resolvable fine structure of the absorption line
for a dipole transition $\tau_{a} \rightarrow \tau_{b}$ will depend on
the relative values of $k\tilde{B}_{a}$, $k\tilde{B}_{b}$, and the
natural linewidth $\Gamma$.  There may be three peaks (if
$k\tilde{B}_{b} \gg \Gamma \gg k\tilde{B}_{b}-k\tilde{B}_{a}$),
$2J_{a}+1$ peaks (if $k\tilde{B}_{b}-k\tilde{B}_{a} \gg \Gamma \gg
k\tilde{B}_{b}$), $3(2J_{a}+1)$ peaks (if $k\tilde{B}_{b}$ and
$k\tilde{B}_{b}-k\tilde{B}_{a} \gg \Gamma$), or more complicated
possibilities when some peaks overlap.  For the $1_{01}\rightarrow 1_{10}$
transition in hydrogen peroxide, our theory predicts that the initial
level spacing is very small, while the final
level spacing is approximately resolvable, for typical
thermal values of $k$.  The expected line shape for this scenario
is schematically depicted in Figure~\ref{fig:fig43}.
There are three peaks, corresponding to the three possible final values of $M$.
The central peak
has the natural linewidth, since the $M=0$ energy has no
first-order dependence on $k$, while the $M=\pm 1$ wings are thermally broadened.

\section{Discussion}
\label{discussion}

In this paper we have presented a classical model for the behavior of
a molecule immersed in superfluid; and we have shown that the
quantized version of the model has interesting features which should
be, under certain conditions, spectroscopically detectable.  In
particular, chiral molecules can act as ``quantum propellers,''
coupling linear and angular momenta via their interaction with the
superfluid medium, and this leads to a characteristic splitting of
spectral lines.  Our model simplifies the superfluid dynamics greatly,
which presumably affects the quantitative accuracy of our results.  We
will conclude by discussing possible remedies for the shortcomings of
our model.

One important physical effect that we have ignored is the formation of
structured shells of helium atoms around a solvated impurity, which
leads to a nonuniform fluid density.  The model can be naturally
extended to include this, by allowing the fluid density $\rho_{f}$ to
vary spatially, while maintaining a time-independent profile in the
body frame.  This approach has been used by other authors to estimate
the superfluid-induced increase in the moments of inertia of linear or
highly symmetric molecules~\cite{kwon,sf6calc,callegariSF}. The effect of a
nonuniform density can be dramatic.  For instance, SF$_{6}$ has $I =
180$\;amu-\AA$^{2}$ in vacuum, which increases by $\Delta I =
310$\;amu-\AA$^{2}$ when the molecule is immersed in a helium droplet. 
The model presented here gives $\Delta I < 25$\;amu-\AA$^{2}$, an
order of magnitude too small, confirming the assessment by other authors
that a rotating ellipsoid model can seriously underestimate $\Delta I$~\cite{sf6calc,callegariSF}.
By contrast, a calculation using a nonuniform density predicts $\Delta I = 170$\;amu-\AA$^{2}$, which is
more than half of the experimental value~\cite{sf6calc}. The comparison
demonstrates that a rigid body with solvation shells attached drags
more mass when it moves than one without them.  It is tempting to
speculate that the calculated chiral splitting constants
$\tilde{B}_{\tau}$ will also increase when a nonuniform density is
allowed; but this may not be the case, for two reasons.  First,
since the off-diagonal Hamiltonian matrix $\hat\beta$ is equal to
$-\hat{M}^{-1}\hat{G}\hat{I}^{-1}$ to lowest order in $\hat{G}$, increases in the
rotational-translational coupling tensor are counteracted by
corresponding increases in the hydrodynamic mass and inertia tensors. 
Second, attaching solvation shells to the immersed body simply
builds a larger rigid body, to some extent, and the constants
$\tilde{B}_{\tau}$ always vanish for a rigid body.  Therefore, while
allowing a nonuniform fluid density may alter our results, it is
difficult to predict the outcome without performing a full
calculation.

By studying the dynamics of immersed molecules in bulk superfluid,
we have ignored the finite-size effects associated with immersion
in a cluster or droplet.  Our approach can be extended to
handle reasonably large droplets with no fundamental difficulties.
Lehmann has carried out a related calculation for the case of a linear
molecule in a spherical droplet, taking into account the hydrodynamic
effects which are present for an elliptical molecular surface (the modification
of the rotational constants and the $\hat\alpha$ matrix, in our notation),
together with the effective interactions between the molecule and the
droplet surface~\cite{lehmann}.  It would be interesting to extend the
current work along similar lines.  We expect that such an extension
would lead to results qualitatively similar to those presented
here.  Because parity is
a good symmetry for a superfluid droplet, as it is for bulk superfluid,
parity would still be broken by the addition of a chiral impurity molecule,
and this would manifest itself in a characteristic splitting of
the rotational spectrum.

Microscopic details of the superfluid structure are also neglected by
a continuum hydrodynamic approach.  These details are better treated
by path-integral and diffusion Monte Carlo methods, which have proven
successful in predicting the rotational constants of immersed
molecules~\cite{kwon}. It is likely that the hydrodynamic mass and
rotational-translational tensors described here can also be extracted
from such calculations, and this information would be a useful
complement to the hydrodynamic results.  Monte Carlo methods
could also be used to address another neglected phenomenon: the
quantum-mechanical tunnelling between left-handed and right-handed
forms of a chiral molecule.  This may play a pronounced role in
hydrogen peroxide, where the torsional ground state (a symmetric
superposition of left- and right-handed forms) is $11$\;cm$^{-1}$
below the first (antisymmetric) excited state~\cite{tunnelsplitting}.
Because the splitting is so large, comparable even to the
rotational level spacing, the enantiomers are strongly mixed,
making questionable our rigid-body treatment of the molecule.
We expect the surrounding superfluid to suppress tunnelling,
but a detailed calculation, without the assumption of rigidity,
is needed for a quantitative assessment.  (Some experimental support
is provided by the work of Nauta and Miller, who studied immersed HF dimers;
they found that the barrier to interchange tunnelling was raised significantly
by the presence of the superfluid~\cite{nauta}.)
The main difficulty in
applying Monte Carlo methods would be finite-size effects: since
translational symmetry is strongly broken in small droplets, while
rotational symmetry is preserved, the finite-size corrections to
$\hat{M}$ and $\hat{G}$ are presumably more drastic than those to
$\hat{I}$.  A number of other microscopic techniques have been applied
to molecular impurities in superfluid droplets (see
Ref.~\cite{kwon} for a recent review), such as density
functional theory, and these more elaborate methods may also be
applicable to the phenomena we have described.

Finally, experimental data would be extraordinarily useful in refining
the current model.  The spectroscopic signatures we have described
should be present for any chiral molecule, but not all candidate
molecules will have the large splitting constants necessary to resolve
the fine structure.  Nanodroplet experiments have been conducted using
many different species of impurity molecule;
Ref.~\cite{HENDIreview} contains an exhaustive list.  However,
almost none of the studied molecules are chiral, so it is not
surprising that no quantum propeller has yet been seen.  We have
suggested hydrogen peroxide (HOOH) as one strong candidate.  Ideally,
the present paper will provide sufficient impetus for more detailed
investigations, both theoretical and experimental.

\begin{acknowledgments}
This work was funded by a Department of Education GAANN Fellowship,
No. P200A970615.
\end{acknowledgments}

\appendix

\section{Calculation of fluid tensors}
\label{sec:appendix}

For approximately spherical surfaces, it is possible to calculate the
hydrodynamic mass, rotational-translational coupling, and inertia
tensors as perturbation series in the deviation from sphericity. 
Consider a surface at $r_{s}(\theta,\phi) =
r_{0}\left(1+\epsilon(\theta,\phi)\right)$.  The surface normal times
the area element is
\begin{equation}
\frac{\nhat da}{r_{0}^{2}d^{2}\Omega} = (1+\epsilon) \left({\bf 
e}_{r} + i \left({\bf e}_{r}\times{\bf J} \right) \right) (1+\epsilon),
\end{equation}
where
\begin{equation}
{\bf J} \equiv i{\bf e}_{\theta}\frac{1}{\sin\theta}
\frac{\partial}{\partial\phi} - i{\bf e}_{\phi}
\frac{\partial}{\partial\theta}
\end{equation}
is the usual vector of first-order differential operators on the 
sphere.  For any scalar function $\Phi$,
\begin{equation}
	\boldgrad\Phi\cdot\frac{\nhat da}{r_{0}^{2}d^{2}\Omega} =
	(1+\epsilon)^{2}\frac{\partial\Phi}{\partial r} + 
	\frac{1}{r_{0}}{\bf J}\Phi\cdot{\bf J}\epsilon,
\end{equation}
with derivatives evaluated at $r=r_{s}$.  Applying this to
Eq.~(\ref{eq:psieqn}) for the translational response function
$\psi_{\mu}$ yields
\begin{equation}
\label{psieqn}
\frac{\partial\psi_{\mu}}{\partial r} = n_{\mu}
-\frac{1}{r_{0}(1+\epsilon)^{2}}{\bf J}\psi_{\mu}\cdot{\bf J}\epsilon
+ \frac{1}{1+\epsilon}
i\epsilon_{\mu\nu\lambda}n_{\nu}({\bf J}\epsilon)_{\lambda},
\end{equation}
where $n_{\mu}={\bf e}_{\mu}\cdot{\bf e}_{r}$, and with derivatives
evaluated at $r=r_{s}$.  If $\psi_{\mu}$ is expanded in orders of 
$\epsilon$, as $\psi_{\mu} = \psi_{\mu}^{(0)} + \psi_{\mu}^{(1)} + 
\ldots$, and functions evaluated at $r_{s}$ are expanded in Taylor
series around $r_{0}$, then Eq.~(\ref{psieqn}) gives a single equation
at each order in $\epsilon$:
\begin{subequations}
\begin{eqnarray}
\frac{\partial\psi_{\mu}^{(0)}}{\partial r} &=& n_{\mu}, \\
\frac{\partial\psi_{\mu}^{(1)}}{\partial r} + 
r_{0}\epsilon\frac{\partial^{2}\psi_{\mu}^{(0)}}{\partial r^{2}} &=&
-\frac{1}{r_{0}}{\bf J}\psi_{\mu}^{(0)}\cdot{\bf J}\epsilon
+i\epsilon_{\mu\nu\lambda}n_{\nu}({\bf J}\epsilon)_{\lambda}, \nonumber\\
\end{eqnarray}
\end{subequations}
and so forth, where now all derivatives are evaluated at $r=r_{0}$.
Moreover, $\grad^{2}\psi_{\mu}^{(n)}=0$ and 
$|\boldgrad\psi_{\mu}^{(n)}({\bf x})| \rightarrow 0$ as $|{\bf x}| 
\rightarrow \infty$ for each $n$.
The zeroth-order equation can be solved for $\psi_{\mu}^{(0)}$, and in general
the equation of order $\epsilon^{n}$ can be solved for $\psi_{\mu}^{(n)}$
once all 
the $\psi_{\mu}^{(m)}$ with $m<n$ are known.  Therefore $\psi_{\mu}$
can be calculated to any order in $\epsilon$, though the process
rapidly becomes tedious.  Similarly, the rotational 
response function $\rotfnc_{\mu}$ satisfies the simpler equation
\begin{equation}
\frac{\partial\rotfnc_{\mu}}{\partial r} = 
-i r_{0} ({\bf J}\epsilon)_{\mu} -\frac{1}{r_{0}(1+\epsilon)^{2}}{\bf J}\rotfnc_{\mu}
\cdot{\bf J}\epsilon
\end{equation}
at $r=r_{s}$, which follows from Eq.~(\ref{eq:rotfnceqn});
and this can be used to find $\rotfnc_{\mu}$ to 
any order in $\epsilon$, in exactly the same way.

Once the
response functions are in hand, Eqs.~(\ref{eqs:fluidtensors}) can
be evaluated to obtain the fluid tensors.  This, too, is done
order-by-order in $\epsilon$.  The results are best expressed as
angular averages.  Through second order in $\epsilon$,
\begin{subequations}
\begin{eqnarray}	
\frac{\delta M_{\mu\nu}}{4\pi\rho_{f}r_{0}^{3}} &=& \frac{1}{6}\delta_{\mu\nu} 
 + \left< \left( -\frac{9}{4}n_{\mu}n_{\nu} + 
\frac{5}{4}\delta_{\mu\nu} \right) \epsilon \right> 
 \nonumber\\ && - \left< \psi_{\mu}^{(1)}\frac{\partial\psi_{\nu}^{(1)}}{\partial r} \right>, \\
\frac{\delta G_{\mu\nu}}{4\pi\rho_{f}r_{0}^{4}} &=& 
\frac{1}{2}\epsilon_{\mu\nu\lambda} \left< n_{\lambda}\epsilon \right> 
 - \left< \psi_{\mu}^{(1)}\frac{\partial\rotfnc_{\nu}^{(1)}}{\partial r} \right>, \\
\frac{\delta I_{\mu\nu}}{4\pi\rho_{f}r_{0}^{5}} &=&
- \left< \rotfnc_{\mu}^{(1)}\frac{\partial\rotfnc_{\nu}^{(1)}}{\partial 
r} \right>,
\end{eqnarray}
\end{subequations}
where $\left<f\right>\equiv\frac{1}{4\pi}\int f(\Omega)d^{2}\Omega$, and functions are 
evaluated at $r_{0}$.  Here the first-order fields $\psi_{\mu}^{(1)}$ and 
$\rotfnc_{\mu}^{(1)}$ are determined by their radial derivatives at $r_{0}$, 
which are
\begin{eqnarray}
\frac{\partial\psi_{\mu}^{(1)}}{\partial r} &=&
-\frac{3}{2}i\epsilon_{\mu\nu\lambda}n_{\nu}({\bf 
J}\epsilon)_{\lambda} - 3 n_{\mu}\epsilon, \nonumber\\
\frac{\partial\rotfnc_{\mu}^{(1)}}{\partial r} &=& -i({\bf 
J}\epsilon)_{\mu}.
\end{eqnarray}
Note that $\delta\hat{G}$ is antisymmetric to first order.  Because
the antisymmetric part of $\hat{G}$ is eliminated by the correct 
choice of body origin, it was necessary to carry the calculation
of $\delta\hat{G}$ to second order to obtain nontrivial 
rotational-translational coupling.

\end{document}